\documentclass[11pt]{article}
\usepackage{latexsym}
\usepackage{amssymb}
\usepackage{graphics}
\usepackage{epsfig}
\usepackage{rotating}
\usepackage{amsfonts}
\usepackage{amsmath}
\usepackage{hyperref}
\hypersetup{
	colorlinks=true,
	linkcolor=blue,
	filecolor=magenta,
	urlcolor=cyan,
}
\begin{document}

\begin{titlepage}
\begin{flushright}
IRMP-CP3-23-10\\
\end{flushright}

\vspace{5pt}

\begin{center}

{\Large\bf The Gravito-Electromagnetic Approximation}

\vspace{5pt}

{\Large\bf to the Gravimagnetic Dipole}

\vspace{5pt}

{\Large\bf and its Velocity Rotation Curve}

\vspace{40pt}

Jan Govaerts\footnote{Fellow of the Stellenbosch Institute for Advanced Study (STIAS), Stellenbosch,
Republic of South Africa}

\vspace{30pt}

{\sl Centre for Cosmology, Particle Physics and Phenomenology (CP3),\\
Institut de Recherche en Math\'ematique et Physique (IRMP),\\
Universit\'e catholique de Louvain (UCLouvain),\\
2, Chemin du Cyclotron, B-1348 Louvain-la-Neuve, Belgium}\\
E-mail: {\em Jan.Govaerts@uclouvain.be}\\
ORCID: {\tt http://orcid.org/0000-0002-8430-5180}\\
\vspace{20pt}

{\sl International Chair in Mathematical Physics and Applications (ICMPA--UNESCO Chair)\\
University of Abomey-Calavi, 072 B.P. 50, Cotonou, Republic of Benin}\\

\vspace{40pt}

\begin{abstract}
\noindent
In view of the observed flat rotation curves of spiral galaxies and motivated by the simple fact that within newtonian gravity a stationary axisymmetric mass distribution
or dark matter vortex of finite extent readily displays a somewhat flattened out velocity rotation curve up to distances comparable to the extent
of such a vortex transverse to the galaxy's disk, the possibility that such a flattening out of rotation curves may rather be a manifestation of some stationary axisymmetric space-time curvature
of purely gravitational character, without the need of some dark matter particles,
is considered in the case of the gravimagnetic dipole carrying opposite NUT charges and in the tensionless limit of its Misner string,
as an exact vacuum solution to Einstein's equations. Aiming for a first assessment of the potential of such a suggestion easier than a full fledged study
of its geodesics, the situation is analysed within the limits of weak field gravito-electromagnetism and nonrelativistic dynamics. Thereby leading indeed
to interesting and encouraging results.

\end{abstract}

\end{center}

\end{titlepage}

\setcounter{footnote}{0}

\section{Introduction}
\label{Intro}

The physical reality of the missing mass, or so-called dark matter problem of the Universe is by now largely accepted and emphasized,
with an accumulation of a diversity and complementary series of converging and convincing indications ranging from the galactic distance scales up to
the astrophysical and cosmological ones {black}{(among many available reviews, see for instance Refs.\cite{DarkWiki,Dark1,Dark2,Strigari})}.
Given the newtonian gravity conceptual paradigm (and Gauss' theorem), it is quite natural to hypothesize that the dark matter conundrum
is the manifestation of the existence of some unknown species of elementary dark matter particles with close to vanishing if any interactions with ordinary matter,
besides their gravitational interactions. For instance the most favoured and studied explanation for the observed flatness of spiral galaxy velocity rotation curves
extending much beyond the visible outer edge of their disks---historically, the first genuine manifestation of a missing mass problem---, involves a (close to)
spherical halo of dark matter particles inside of which sits the visible rotating galaxy \textcolor{black}{(among the vast literature on the subject, see for
instance Refs.\cite{Sofue-Rubin,Sofue1,Sofue2,Sofue3,Zatri1,Zatri2}, and references therein)}.
Indeed the dark matter particle paradigm appears nowadays to be widely accepted by most to necessarily provide the actual {\sl raison d'\^etre\/}
of the indirectly observed missing mass mystery of the Universe at all scales. One just needs to identify which new type(s) of species of particles it may well be,
and this by any means imaginable and available.

Yet, in spite of many decades now of much genuine physical and experimental ingenuity and considerable instrumental efforts,
and in parallel much theoretical inventiveness,
to this day there does not exist any experimentally established direct evidence for the existence of such dark matter particles.
Which is a reason presumably why possible alternatives
discarded by most in earlier days, are now being studied anew, in earnest and with an increased sense of possible relevance. Such alternatives
are being addressed whether by remaining within the conceptual paradigms of newtonian gravity or of General relativity, or beyond these through
modifications, generalisations or deformations thereof.

Consider thus for example the situation of the close to flat velocity rotation curves of spiral galaxies.\footnote{\textcolor{black}{Because of the nonvanishing
conserved angular-momentum sourced by the doublet of rotating NUT black holes of the exact gravimagnetic dipole solution to General relativity
to be considered in its first astrophysical exploration discussed herein, one has specifically the case
of spiral galaxies in mind as a situation of comparison which obviously also possess such an angular-momentum sourced by some rotating matter distribution.
However this does not necessarily preclude some eventual relevance of the gravimagnetic dipole solution to the case of elliptical and spheroidal galaxies as well,
which are known to also share in the missing mass mystery \cite{Strigari}.}}
Within newtonian gravity such motion is to follow from
an axisymmetric gravitational field sourced by some stationary axisymmetric mass distribution. Indeed the velocity rotation curve (for circular
trajectories within the galaxy's disk) then follows from the expression (see for instance the Appendix)
\begin{equation}
v(r)=\sqrt{r\frac{\partial \phi(r,\cos\theta=0)}{\partial r}},
\end{equation}
where $\phi(r,\cos\theta)$ is the newtonian gravitational scalar potential sourced by the mass distribution, and $(r,\theta,\varphi)$ are the
spherical coordinates relative to the rotational symmetry axis $z=r\cos\theta$ of the mass distribution. For a close to spherically symmetric mass distribution
of visible matter confined to within a radius $R$, and because of Gauss' theorem, starting at distances $r$ not much larger than $R$ the rotation curve
should essentially follow already a leading $1/\sqrt{r}$ overall behaviour, of course different from its observed flatness---the gravitational pull provided
by the visible mass distribution thus proves to be too weak at the larger distances. A natural and simple hypothesis is obviously to suggest
that the missing mass may be provided by a spherical halo of dark matter particles of radius much larger than $R$.

On the other hand however, if $v(r)$ is to remain essentially constant within a large interval of $r$ values (relative to the $R$), the above relation
implies that the gravitational potential should display essentially a ($\ln r$) behaviour in that region. Of course, given our
\textcolor{black}{undergraduate physics} knowledge of the electric field and potential
sourced by a linear uniform charge distribution of infinite extent, this observation is a direct invitation to suggest \cite{Zatri1,Zatri2,Slovick,Govaerts,Llanes}
that the invisible axisymmetric mass distribution which accounts for the flatness of the rotation curve is aligned along the rotation symmetry axis
of the spiral galaxy, and is of large finite extent along that direction.

Indeed taking a cylinder of total length (or height) $2L$, of radius $a$, and of total mass $2m$ with a uniform
volumetric mass distribution, to a excellent degree of approximation the exact velocity rotation curve of that gravitational system outside the cylinder
is represented \cite{Govaerts} by the following radial profile (see the Appendix),
\begin{equation}
r>a\,:\qquad
v(r)\simeq \sqrt{G\cdot\frac{2m}{L}}\cdot\frac{1}{\left(1+r^2/L^2\right)^{1/4}}=\sqrt{G\cdot\frac{2m}{r}}\cdot\frac{1}{\left(1+L^2/r^2\right)^{1/4}},
\end{equation}
$G\simeq 4.3\times 10^{-6}\,{\rm kpc}\cdot M^{-1}_\odot\cdot\left({\rm km/s}\right)^2$ being Newton's constant.
The net effect of the modulation factor $(1+L^2/r^2)^{-1/4}$ which multiplies the large distance profile $\sqrt{G\cdot 2m/r}$
is indeed to flatten out the rotation curve in the region $a<r<L$ and even beyond the distance $r=L$ before the $1/\sqrt{r}$ behaviour sets in,
as if the flat portion of the rotation curve were a ``mirror image'' of the transverse extent of the axisymmetric mass distribution of some invisible
dark matter threading the centre of the spiral galaxy, which may well not be in the form of dark matter particles \cite{Govaerts}. And thus without the necessary
need otherwise of a spherical halo of dark matter particles of radius much larger than that of the visible edge of the galaxy disk to account for
the possibility of a flat rotation curve. Nonetheless, the suggestion of a dark matter vortex threading the centre of the galaxy directly raises
the remaining open question of the possible physical nature of such form of matter.

Alternatively and now within the context of General relativity, one may entertain the idea that missing mass observations in some given spatial
region are the manifestation of a space-time curvature in that region sourced by some yet unobserved energy-momentum distribution (possibly of
some yet unknown form of matter) localised in some other region of space, hence corresponding to some solution to the vacuum Einstein equations
outside the latter spatial region. Once again more specifically in the case of spiral galaxies and their rotation curves, stationary axisymmetric solutions
could be of relevance. As a matter of fact, dating back to the very first days of General relativity with Weyl's construction of exact static axisymmetric solutions
to Einstein's vacuum equations \cite{Weyl,Weyl2}, and extended over the decades since, large classes of exact stationary axisymmetric solutions
to the same equations are now available \cite{ExactGR}, through powerful techniques of integrable systems applied to the nonlinear Einstein equations.
In view of the above observation of the possible interest of compact axisymmetric mass distributions with large transverse extent as compared
to the galaxy's disk radius, one such General relativity vacuum solution stands out as a case of choice towards a first exploration of such ideas,
namely the so-called ``gravimagnetic dipole'' \cite{Clement,Manko}.

To describe it in simple terms, this stationary solution to the vacuum Einstein equations is sourced by two rotating black holes of identical
masses $m$---their gravi-electric ``monopole charges''---but also carrying opposite valued NUT (Newman-Unti-Tamburino)
charges $\pm \nu$---their gravi-magnetic ``monopole charges''---,
thus connected by a Misner string singularity \textcolor{black}{\cite{Misner1,Misner2,Bossard,Clement2,Awad}}
for the space-time metric---\textcolor{black}{namely nothing other than} the gravimagnetic analogue of the \textcolor{black}{singular magnetic flux carrying}
Dirac string for a Dirac monopole \cite{Dirac1}.
The analogue electromagnetic configuration is that of two dyons of identical monopole electric charges $q$ and opposite monopole magnetic charges $\pm g$.
\textcolor{black}{In that latter case the Dirac string then connects} the two dyons \textcolor{black}{(without any extension to infinity), as does
the gravimagnetic flux carrying Misner string for the gravimagnetic dipole solution with its two black holes of identical masses $m$ and opposite
NUT charges $\pm \nu$}. Furthermore when $|q|=|g|$ (in natural electromagnetic units) the dyons remain
in static equilibrium whatever their relative spatial position and distance. In a likewise manner, for the gravimagnetic dipole the Misner string becomes tensionless
with the two black holes then remaining in static equilibrium at a specific distance, when their mass and NUT charge values meet a specific condition \cite{Clement}.

By analogy with the situation outlined above involving a dark matter vortex within the context of newtonian gravity, such gravimagnetic dipole
space-time metrics could well provide a new vista from which to consider the issue of flat rotation curves for spiral galaxies within the context of General
relativity. With the Misner string then playing a role akin to that of some dark matter axisymmetric vortex but now without the need of any form of matter
to be directly involved, but rather as a (purely gravitational and thus) ``dark'' manifestation of some specific continuously distributed axisymmetric
and (close to) singular structure within the space-time metric, sourced by some largely delocalised energy-momentum distribution.

A thorough assessment of the potential offered by such a solution to Einstein's equations
would require a full fledged (and numerically involved) analysis of the geodesics of the gravimagnetic
dipole space-time, which is not the purpose of the present discussion. Rather, and as a first exploration of the possible viability of the above suggestions
with regards to the dark matter mystery, the \textcolor{black}{asymptotic} weak field, or gravito-electromagnetic approximation of that space-time metric
is addressed. With the aim of
identifying then the rotation curve for nonrelativistic circular trajectories within that space-time geometry, and assess whether it could improve sufficiently
on the large distance $1/\sqrt{r}$ behaviour implied by newtonian gravity to eventually allow for sufficiently flattened out rotation curves before that
$1/\sqrt{r}$ behaviour sets in, and thereby become comparable to what is being observed \textcolor{black}{in the case, for example, of} spiral galaxies.

\textcolor{black}{
The analysis presented herein is constructed along the following considerations. Albeit rather involved, the gravimagnetic dipole solution is known in terms of an exact
analytic expression for its space-time metric involving three physical parameters. Since this metric is asymptotically flat, its weak field regime is reached
at large distances, thereby allowing for approximate representations through power series in $1/r$, with the distance $r$ measured relative to some distance scale characteristic of the solution. As an exact solution to Einstein's field equations, the weak field regime is a ready solution to the linearised Einstein equations,
and provides as well a linearised geodesic equation for trajectories of classical point particles sufficiently distant from the gravimagnetic dipole.
Within such a linearised approximation, the equations of General relativity may be brought to a form analogous to that of Maxwell's equations
of classical electrodynamics coupled to the Lorentz force for classical point charges, known as gravito-electrodynamics.
In the present case since the space-time metric is known already as an exact solution, there is no need to consider the dynamical equations
for the gravito-electromagnetic fields. One may concentrate directly on the geodesic equation in the form of a Lorentz-like force involving
the gravito-electromagnetic fields. And thereby restrict to circular trajectories in the plane transverse to the
gravimagnetic dipole symmetry axis, thus eventually determining the velocity rotation curve of gravimagnetic dipole in the nonrelativistic and large distance
approximation.}

\textcolor{black}{
Equations for gravito-electrodynamics have a long history, in a way going back even to before when General relativity was conceived \cite{Williams}. The gravito-electrodynamic
Lorentz-like force equation used in the present work is in the form obtained specifically in Ref.\cite{Williams}. Through a careful and detailed analysis, that reference
assesses the consequences for the linearised Einstein field equations of the freedom in the choice of gravitational gauge fixing condition in relation to their symmetry
under space-time coordinate diffeomorphisms. As established in Ref.\cite{Williams}, the form
of the gravito-electromagnetic field equations are dependent on that gauge choice. That same reference also points out the importance for these equations
to include an extra scalar field, called the {\sl neutral field} in addition to the gravito-electric and -magnetic fields, which as a matter of fact also couples
to the gravito-electrodynamic Lorentz-like force equation. These features are too often overlooked in the literature.
Quite often for an implementation of gravito-electrodynamics one relies on the discussion presented in Refs.\cite{Mash1,Mash2} which however,
do not properly account for such more subtle aspects, and do not necessarily present a self-consistent linearisation of the geodesic equation
in particular in the nonrelativistic regime, in contradistinction to Ref.\cite{Williams}.}

\textcolor{black}{
Calling on gravito-electrodynamics rather than haloes of dark matter particles
in trying to account for flat rotation curves of (spiral) galaxies in particular, is by no means a new idea.
Different attempts and detailed analyses based on a variety of empirical models for the visible mass distributions of specific galaxies, without including
any other invisible or dark matter contribution but rather by accounting for the gravito-magnetic effects sourced by visible mass currents,
have been contributed over the years in a number of publications, such as Refs.\cite{Barros,LeCorre,Nyam,Ludwig,Toth,Ciotti,Ruggiero,Astesiano1,Astesiano2}
(and references therein). However, these authors come to dissimilar conclusions---some (strongly) supporting the possibility, while others being not (at all)
as convinced, and some others still even discouraged of pursuing such an avenue---, depending on the models being used for visible matter distributions
and their dynamics. It appears that a final verdict regarding that suggestion remains to be reached. For instance in the recent interesting
work of Ref.\cite{Ludwig}, by adjusting a fair number of free parameters used to model visible mass distributions and their dynamics
for a number of galaxies, the author manages to fit their measured rotation curves including their plateau
by including in the dynamical equations the gravito-magnetic corrections sourced by the visible mass currents,
thus without any extra dark matter contribution. In such an approach one needs to solve both the gravito-electromagnetic field
equations for the gravitational fields sourced by an {\sl ad hoc}  model for the visible matter distribution and its currents,
as well as the coupled Lorentz-like force equation for the dynamics of that same visible matter distribution.}

\textcolor{black}{
The approach implemented in the present analysis differs certainly from those having just been mentioned in at least three main respects, besides the fact
that the formulation of gravito-electrodynamics being used is that specifically carefully developed in Ref.\cite{Williams} (in contradistinction to other works). First,
certainly no visible matter distribution whatsoever is involved, but rather solely the gravimagnetic dipole with its two rotating black holes of identical masses
and opposite NUT charges, connected by their Misner string, as an exact strong field solution to Einstein's field equations with its string-like
space-time singularity structure. Second, our discussion relies
solely on the gravito-electromagnetic Lorentz-like force equation, while not at all considering the gravito-electromagnetic field equations for the
gravitational field since the latter are guaranteed to be met for the simple fact that the space-time metric being used is already an exactly known nonperturbative
or nonlinear solution to the full nonlinear Einstein field equations. Thirdly, given the mass scale $m$ associated to that solution, it involves only two
extra parameters, namely the distance between the two black holes and the value of their opposite NUT charges, which as a matter of fact reduce to a single
independent free parameter in the tensionless limit of their Misner string. In effect, since the mass scale also sets the distance scale for the gravitational interaction
in a relativistic setting through the factor $Gm/c^2$ possessing a physical dimension of length, the tensionless
gravimagnetic dipole solution is characterised by a single free 
and dimensionless parameter, the value of which remains to be chosen in order to lead not only to a sufficiently wide flat plateau in the velocity curve, but as well
to a value of the plateau velocity which compares well with those values that are indeed observed. And this in the absence still of the additional contributions
of some visible matter distribution and its mass current density. Even though reached within the weak field and nonrelativistic approximations being implemented,
as the main conclusion of the present work it will be established that both these objectives are readily met with an appropriate choice of value
for that single free parameter. And this already without the need---necessary in all other approaches---to adjust the profile of whatever visible mass distribution
that ought to be included (at a later stage) towards more physically realistic assessments of the possible relevance of gravimagnetic dipoles
and their role in attempting an understanding of the dark matter mystery.}

The discussion hereafter is \textcolor{black}{hence}
structured as follows\textcolor{black}{, and thus implements the general programme outlined above}.
In Section 2, the space-time metric of the gravimagnetic dipole is considered, of which the $1/r$ asymptotic
expansion is constructed up to the order required for our purposes in Section 3. Section 4 then identifies the relevant gravito-electromagnetic fields \cite{Williams}
for that vacuum solution to Einstein's equations---in Subsection~4.1---, to finally identify a resummed representation of the corresponding velocity rotation curve
in the nonrelativistic regime within that same gravito-electromagnetic approximation, with a first numerical assessment of its physical potential to account
for the observed flattened out component of the rotation curve of spiral galaxies. Some final comments are then presented in the Conclusions.
While further considerations and relations of use and reference in the main text \textcolor{black}{relating to newtonian gravity are presented in an Appendix.}

\section{The  Space-Time Metric of the Gravimagnetic Dipole}
\label{Sect2}

Let us consider the gravimagnetic dipole solution to the vacuum Einstein equations in the notations and conventions
of Refs.\cite{Clement,Manko}, with their implicit use of natural units such that\footnote{We thank G\'erard Cl\'ement \cite{Clement}
for confirming this choice of units in a private e-mail communication.} $c=1$ and $G=1$. This configuration consists of the nonlinear superposition
of two NUT objects of equal masses $m> 0$ but opposite NUT charges $\pm \nu$ (without loss of generality one may take $\nu\ge 0$, as assumed
throughout hereafter), separated by a total distance $2k\ge 2m>0$, and aligned symmetrically around $z=0$ along the $z$ coordinate axis.

In Weyl coordinates $(t,\rho,\varphi,z)$ the corresponding stationary asymptotically flat space-time metric is of the form,
\begin{equation}
ds^2=-f\left(dt - \omega d\varphi\right)^2 + f^{-1}\left[e^{2\gamma}\left(d\rho^2+dz^2\right)+\rho^2d\varphi^2\right],
\label{eq:metric}
\end{equation}
where $f(\rho,z)$, $\gamma(\rho,z)$ and $\omega(\rho,z)$ are functions only of $(\rho,z)$. These functions are constructed as follows from complex valued
potentials $A$, $B$ and $G=G_2+G_1$ \textcolor{black}{to be detailed presently} (which determine the Ernst potential for the solution \cite{Clement,Manko}),
\begin{equation}
f=\frac{|A|^2-|B|^2}{|A+B|^2},\qquad
e^{2\gamma}=\frac{|A|^2-|B|^2}{64\, d^4\, \alpha^2_+ \alpha^2_- \, R_+ R_- r_+ r_-},\qquad
\omega=-4\,\frac{{\rm Im}\left[G(\bar{A}+\bar{B})\right]}{|A|^2-|B|^2}.
\label{eq:AB}
\end{equation}
The explicit expressions for these potentials $A$, $B$, $G_1$ and $G_2$ \textcolor{black}{are
listed hereafter} as functions of $(\rho,z)$ and \textcolor{black}{involve in particular the specific quantities}
\begin{equation}
R_\pm=\sqrt{\rho^2+(z\pm\alpha_+)^2},\qquad
r_\pm=\sqrt{\rho^2+(z\pm\alpha_-)^2},
\label{eq:Rrpm}
\end{equation}
where
\begin{equation}
\alpha_\pm=\sqrt{m^2+k^2-\nu^2\pm 2 d},\qquad
d=\sqrt{m^2 k^2 + \nu^2(k^2-m^2)}.
\label{eq:alphapm}
\end{equation}

\textcolor{black}{
Refs.\cite{Clement,Manko} provide the following expressions for the complex valued potentials $A$, $B$, $G_1$ and $G_2$ in terms of which
the parametrisation of the space-time metric of the gravimagnetic dipole as given in (\ref{eq:metric}) is constructed.
For the $A(\rho,z)$ potential one has,
\begin{eqnarray}
A &=& \left[(m^2+\nu^2)(k^2-m^2)(k^2-m^2-\nu^2) - 2 m^2 k^2 \nu^2\right]\,(R_+ - R_-)(r_+ - r_-) \nonumber \\
&& -\, \alpha_+ \alpha_- \left[\quad 2(m^2+\nu^2)(k^2-m^2)(R_+ R_- + r_+ r_-) \right. \nonumber \\
&&\qquad\qquad \left. + (2m^4+(m^2+\nu^2)(k^2-m^2))(R_+ + R_-)(r_+ + r_-)\right] \nonumber \\
&&-\, 2im k\nu d\left[(\alpha_+ - \alpha_-)(R_+ r_+ -R_- r_-) - (\alpha_+ + \alpha_-)(R_+ r_- -R_- r_+)\right],
\label{eq:A}
\end{eqnarray}
while the $B(\rho,z)$ potential reads,
\begin{eqnarray}
B &=& \quad 4d\,m\alpha_+ \alpha_- \left[(m^2-d)(R_+ + R_-) - (m^2+d)(r_+ + r_-)\right] \nonumber \\
&&+ i\, 4 k\nu d \left[\alpha_-(m^2-d)(R_+ - R_-) - \alpha_+ (m^2+d) (r_+ - r_-)\right].
\label{eq:B}
\end{eqnarray}
Finally the $G_2(\rho,z)$ and $G_1(\rho,z)$ potentials are given as, respectively,
\begin{eqnarray}
G_2 &=& -d\left[d^2+m^2(m^2+2i k\nu)\right]\left[(\alpha_+ - \alpha_-)(R_+ r_+ -R_- r_-) - (\alpha_+ + \alpha_-)(R_+ r_- - R_- r_+)\right] \nonumber \\
&& + 2 m^2 d^2\left[(\alpha_+ + \alpha_-)(R_+ r_+ - R_- r_-)-(\alpha_+ - \alpha_-)(R_+ r_- - R_- r_+)\right] \nonumber \\
&& - m \alpha_+ \alpha_- (d^2+m^4)(R_+ + R_-)(r_+ +r_-) \nonumber \\
&&+ m\left[kd^2(k+4i\nu) - (2k^2-m^2)(m^2+\nu^2)^2+k^2\nu^4\right] (R_+ - R_-)(r_+ - r_-) \nonumber \\
&&-2m\alpha_+ \alpha_- (m^2+\nu^2)(k^2-m^2)(R_+ R_- + r_+ r_-) \nonumber \\
&&-2d\,z\left\{\ \  \alpha_-(m^2-d)\left[m\alpha_+(R_+ +R_-) + i\,k\nu(R_+ - R_-)\right] \right.  \nonumber \\
&& \qquad\quad \left. - \alpha_+ (m^2+d) \left[ m\alpha_- (r_+ + r_-) + i\,k\nu (r_+ - r_-)\right]\right\},
\label{eq:G2}
\end{eqnarray}
and
\begin{eqnarray}
G_1 &=& 2 d \alpha_+ \alpha_-(2m^2+i\,k\nu)\left[m^2(R_+ + R_- - r_+ - r_-) - d(R_+ + R_- + r_+ +r_-)\right] \nonumber \\
&& - 2 md\left[d^2-m^4-i\,k\nu(2m^2+i\,k\nu)\right]\left[\alpha_-(R_+ - R_-) - \alpha_+(r_+ - r_-)\right] \nonumber \\
&&+ 2 md^2(m^2-k^2+\nu^2-2\,i\,k\nu)\left[\alpha_-(R_+ - R_-) + \alpha_+(r_+ - r_-)\right].
\label{eq:G1}
\end{eqnarray}
}

For a detailed discussion of the main physical properties of this solution to the vacuum Einstein equations the reader is referred to Ref.\cite{Clement}.
In particular this asymptotically flat metric possesses a total mass $M=2m$ and a total angular momentum $J=2k\nu$.
Note that given the choice of natural units, all distance scales may be expressed in units of $m$.
For sufficiently large values of $k\ge m$, the system consists of two black holes connected by a spinning Misner
string \textcolor{black}{\cite{Misner1,Misner2,Bossard,Clement2,Awad}}. The quantities $\alpha_-<\alpha_+$
and $-\alpha_+<-\alpha_-$ correspond to the positions of the intercepts of the horizons of these black holes on the $z$-axisymmetric axis,
while the Misner string lies along $-\alpha_- < z < \alpha_-$. \textcolor{black}{However, a few more explicit general considerations are in order here
so as to prepare a series of tools of relevance to the analysis presented in the remainder of the paper,
beginning specifically with Sect.\ref{Sect3} and the weak field asymptotic expansion of the above space-time metric.}

\textcolor{black}{
When evaluating the weak field asymptotic expansion, given a value for $m$}
rather than using $(\nu,k)$ as two further independent parameters it proves efficient to \textcolor{black}{effect a change of parametrisation and} work
in terms of the following two alternative dimensionless quantities \cite{Clement} representing the mean and difference values of $\alpha_+$ and $\alpha_-$ 
relative to $M=2m$,
\begin{equation}
\sigma=\frac{\alpha_+ +\alpha_-}{2m},\qquad
\delta=\frac{\alpha_+ - \alpha_-}{2m},\qquad
\sigma\ge\delta > 0.
\end{equation}
One then has the \textcolor{black}{inverse} relations,
\begin{equation}
d=m^2\sigma\delta,\qquad
k^2\nu^2=m^4(\sigma^2-1)(\delta^2-1),\qquad
k^2-\nu^2=m^2(\sigma^2+\delta^2-1).
\end{equation}
Note that the property $k^2\nu^2\ge 0$ requires the necessary conditions that $\sigma \ge \delta\ge 1$.
In addition it follows that the dimensionless parameters $k/m\ge 1$ and $\nu/m\ge 0$ are then given as,
\begin{equation}
\frac{k^2}{m^2}=\frac{1}{2}\left[\sqrt{(\sigma^2+\delta^2-1)^2+4(\sigma^2-1)(\delta^2-1)} \ + \ (\sigma^2+\delta^2-1)\right],
\end{equation}
\begin{equation}
\frac{\nu^2}{m^2}=\frac{1}{2}\left[\sqrt{(\sigma^2+\delta^2-1)^2+4(\sigma^2-1)(\delta^2-1)} \ - \ (\sigma^2+\delta^2-1)\right].
\end{equation}

Finally \textcolor{black}{when coming to the final expressions in Sect.\ref{Sect4} relevant to the gravito-electromagnetic fields to be associated
to the weak field approximation established in Sect.\ref{Sect3},
yet another change of parametrisation proves to be most useful. To this aim}, consider the usual quantity representing
the total angular momentum $J$ per total mass $M$, namely $a=|J|/M$, as well as the dimensionless quantity $\xi\ge 0$ defined as
\begin{equation}
\xi=\frac{a}{M}=\frac{|J|}{M^2}=\frac{1}{2}\,\frac{k|\nu|}{m^2}=\frac{1}{2}\sqrt{(\sigma^2-1)(\delta^2-1)}\ge 0.
\end{equation}
\textcolor{black}{Given a value for the mass $m$}, this parameter $\xi$ allows for still another double dimensionless
parametrisation \textcolor{black}{of the gravimagnetic dipole solution. This change of variable involves another pair of quantities denoted}
$(\alpha\ge 0,\xi\ge 0)$ \textcolor{black}{and constructed in terms
of $(\sigma,\delta)$,  with $\xi$ given by the expression above while the parameter $\alpha\ge 0$ is defined as,}
\begin{equation}
\alpha=\frac{1}{4}\left(\sigma^2+\delta^2-2\right),\qquad
\sigma^2+\delta^2-2=4\alpha=\frac{k^2-\nu^2}{m^2}-1\ge 0.
\end{equation}
\textcolor{black}{The inverse relations for $(\nu,k)$ then read,}
\begin{equation}
\frac{k^2}{m^2}=\frac{1}{2}\left[\sqrt{(4\alpha+1)^2+16\xi^2} \ + \ (4\alpha+1)\right],
\end{equation}
\begin{equation}
\frac{\nu^2}{m^2}=\frac{1}{2}\left[\sqrt{(4\alpha+1)^2+16\xi^2} \ - \ (4\alpha+1)\right].
\end{equation}

\textcolor{black}{In addition to} these general considerations, let us remark that the particular situation of interest when no NUT charge is involved,
namely the double black hole solution with $\nu=0$, is characterised by the values of $\alpha=(k^2/m^2-1)/4\ge 0$ and $\xi=0$, or equivalently,
\begin{equation}
d=mk,\quad
\alpha_\pm=k\pm m\ge 0,\quad
\sigma=\frac{k}{m}\ge 1,\quad
\delta=1,\quad
\frac{k^2}{m^2}=\sigma^2\ge 1,\quad
\frac{\nu^2}{m^2}=0,
\end{equation}
with thus a distance $2k=2m\sqrt{4\alpha+1}=2m\sigma\ge 2m$ between the two black holes that remains arbitrary.
In other words, in the region $\sigma^2\ge\delta^2\ge 1$
in the $(\sigma^2,\delta^2)$ plane, the line $\delta^2=1$ with $\sigma^2=k^2/m^2\ge 1$ is that of the solutions with $\nu=0$.
Or equivalently in the quadrant $(\alpha\ge 0,\xi\ge 0)$ in the $(\alpha,\xi)$ plane, it is the line $\xi=0$ with $\alpha\ge 0$
which is that of those same solutions with $\nu=0$.

As is discussed in Ref.\cite{Clement}, \textcolor{black}{when $\nu\ne 0$ and}
as a function of the distance $k\ge m$ the tension in the singular Misner string measures the force which is necessary
to balance the gravitational force between the two sources, attractive between the two equal masses and repulsive between the \textcolor{black}{two}
opposite NUT charges.
Hence the Misner string may become tensionless, with then a balanced configuration, which is a situation of noteworthy physical relevance
that applies when the following restriction between the values for $\sigma$ and $\delta$ is met \cite{Clement},
\begin{equation}
\delta^2=2-\frac{1}{\sigma^2}.
\end{equation}
Correspondingly, since in that case $(\sigma^2-1)(\delta^2-1)=(\sigma^2-1)(1-\sigma^{-2})=(\sigma-\sigma^{-1})^2$, one finds,
\begin{equation}
\xi=\cfrac{1}{2}\left(\sigma - \frac{1}{\sigma}\right)\ge 0.
\end{equation}
In other words, in terms of the then only remaining free parameter $\xi\ge 0$ in addition to $m\ge 0$, tensionless configurations
are characterised by the following values for $\sigma$, $\delta$, and $\alpha$,
\begin{equation}
\sigma^2=1+2\xi\left(\sqrt{\xi^2+1}+\xi\right),\qquad
\sigma=\sqrt{\xi^2+1} + \xi,\qquad
\frac{1}{\sigma}=\sqrt{\xi^2+1} - \xi,
\end{equation}
\begin{equation}
\delta^2=1+2\xi\left(\sqrt{\xi^2+1}-\xi\right),\qquad
\alpha=\xi\sqrt{\xi^2+1},
\end{equation}
which in turn imply \textcolor{black}{the inverse relations},
\begin{equation}
\frac{k^2}{m^2}=\frac{1}{2}\left[\sqrt{\left(1+4\xi\sqrt{\xi^2+1}\right)^2+16\xi^2} \,+\,\left(1+4\xi\sqrt{\xi^2+1}\right)\right],
\end{equation}
\begin{equation}
\frac{\nu^2}{m^2}=\frac{1}{2}\left[\sqrt{\left(1+4\xi\sqrt{\xi^2+1}\right)^2+16\xi^2} \,-\,\left(1+4\xi\sqrt{\xi^2+1}\right)\right].
\end{equation}
Indeed these expressions are such that,
\begin{equation}
\frac{k^2-\nu^2}{m^2}=\sigma^2+\delta^2-1=1+4\xi\sqrt{\xi^2+1},\qquad
\frac{k^2 \nu^2}{m^4}=(\sigma^2-1)(\delta^2-1)=4\xi^2,
\end{equation}
as it should.

In the $(\sigma^2,\delta^2)$ plane the line $(\sigma^2\ge 1,\delta^2=2-1/\sigma^2\ge1)$ of tensionless configurations
meets the line $(\sigma^2\ge 1,\delta^2=1)$ of solutions with $\nu=0$ specifically only for the value $\xi=0$, in which case one has,
\begin{equation}
{\rm Tensionless\ with}\ \xi=0:\quad \sigma^2=1,\quad \delta^2=1,\quad \alpha=0,\quad \frac{k^2}{m^2}=1,\quad \frac{\nu^2}{m^2}=0.
\end{equation}

Furthermore, \textcolor{black}{still for tensionless configurations and of use when considering numerical evaluations of the analytic approximations
established in this work, note the following small or large $\xi$ representations of these different relations.}
For small values of $\xi$ one has following series approximations,
\begin{eqnarray}
\sigma^2 &=& 1+2\xi+2\xi^2+\xi^3-\frac{1}{4}\xi^5+{\cal O}(\xi^7), \nonumber \\
\delta^2 &=& 1 + 2\xi - 2\xi^2 +\xi^3-\frac{1}{4}\xi^5+{\cal O}(\xi^7), \nonumber \\
\alpha &=& \xi + \frac{1}{2}\xi^3-\frac{1}{8}\xi^5+{\cal O}(\xi^7), \nonumber \\
\frac{k^2}{m^2} &=& 1 + 4\xi + 4\xi^2-14 \xi^3 +{\cal O}(\xi^4), \nonumber \\
\frac{\nu^2}{m^2} &=& 4\xi^2\left(1-4\xi+12\xi^2-18\xi^3+{\cal O}(\xi^4)\right).
\end{eqnarray}

As it turns out in the other limit with $\xi\rightarrow\infty$, for tensionless configurations and sufficiently large distances $2k$ between the two NUT black holes,
the parameter $4\xi$ essentially measures, in units of $m$, the total distance between the two objects, while their NUT charges then
coincide essentially with their mass $m$. Indeed one has the following asymptotic dependencies for large values of $\xi$,
\begin{equation}
\xi\rightarrow\infty:\ \ 
\frac{k^2}{m^2}\simeq 4\xi^2\left(1+\frac{1}{\xi^2} -\frac{3}{8}\cdot\frac{1}{\xi^4}+{\cal O}(1/\xi^6)\right),\ \ 
\frac{\nu^2}{m^2}\simeq 1 - \frac{1}{\xi^2}+\frac{11}{8}\cdot\frac{1}{\xi^4}+{\cal O}(1/\xi^6),
\end{equation}
corresponding to,
\begin{equation}
\xi\rightarrow\infty:\ \ 
\sigma^2\simeq 4\xi^2\left(1+\frac{1}{2}\cdot\frac{1}{\xi^2}-\frac{1}{16}\cdot\frac{1}{\xi^4}{+\cal O}(1/\xi^6)\right),\ \
\delta^2\simeq 2\left(1-\frac{1}{8}\cdot\frac{1}{\xi^2}+\frac{1}{16}\cdot\frac{1}{\xi^4}+ {\cal O}(1/\xi^6)\right),
\end{equation}
as well as,
\begin{equation}
\xi\rightarrow\infty:\ \ 
\alpha\simeq \xi^2\left(1+\frac{1}{2}\cdot\frac{1}{\xi^2}-\frac{1}{8}\cdot\frac{1}{\xi^4}+{\cal O}(1/\xi^6)\right).
\end{equation}

\section{The Weak Field Asymptotic Expansion of the Metric}
\label{Sect3}

Given the space-like Weyl coordinates $(\rho,\varphi,z)$, let us introduce the associated radial variable $r$
as well as the spherical angle $\theta$ relative to the $z$ axis, such that,
\begin{equation}
r=\sqrt{\rho^2+z^2},\qquad
z=r\,\cos\theta.
\end{equation}
The coordinates $(r,\theta,\varphi)$ may then be viewed as spherical coordinates relative to the $z$ axis and centered onto $z=0$.

The asymptotic expansion of the gravimagnetic dipole space-time metric for large values of $r$ is then to be effected in terms of series
expansions in the dimensionless variable
\begin{equation}
u=\frac{m}{r}.
\end{equation}
\textcolor{black}{As is established in Sect.\ref{Sect4}, when considering the gravito-electromagnetic fields
and their contributions to the velocity curve in the weak field approximation it turns out that these expansions in $u=r/m$ need}
to be computed exactly up to order $u^4$ inclusive, for the potentials $A$ and $B$,
as well as for the distance factors $R_\pm$ and $r_\pm$ introduced \textcolor{black}{in (\ref{eq:Rrpm}), as done hereafter}.
For example given the scaling in $r$
of the latter quantities, the relevant factors of $r$ need to be factored out properly, leading to the following expansions
for these distance factors, as they contribute to the potentials $A$, $B$ and $G$, exact up to order $u^4$,
\begin{eqnarray}
\frac{1}{r}\,R_\pm &=& 1\pm(\sigma + \delta)\,\cos\theta\cdot u + \frac{1}{2}(\sigma+\delta)^2\,\sin^2\theta\cdot u^2 \nonumber \\
&& \ \mp\,\frac{1}{2}(\sigma+\delta)^3\,\sin^2\theta\,\cos\theta\cdot u^3 \,-\,\frac{1}{8}(\sigma+\delta)^4\,\sin^2\theta\,(1-5\cos^2\theta)\cdot u^4\,+\,{\cal O}(u^5),
\end{eqnarray}
\begin{eqnarray}
\frac{1}{r}\,r_\pm &=& 1\pm(\sigma - \delta)\,\cos\theta\cdot u + \frac{1}{2}(\sigma-\delta)^2\,\sin^2\theta\cdot u^2 \nonumber \\
&&\ \mp\,\frac{1}{2}(\sigma-\delta)^3\,\sin^2\theta\,\cos\theta\cdot u^3 \,-\,\frac{1}{8}(\sigma-\delta)^4\,\sin^2\theta\,(1-5\cos^2\theta)\cdot u^4\,+\,{\cal O}(u^5).
\end{eqnarray}
\textcolor{black}{It should be emphasized} that the tensionless condition is not required when working out the consequences of the asymptotic expansion
for the space-time metric in the present section. The parameters $\sigma\ge\delta\ge 1$, or $(\alpha\ge 0,\xi\ge 0)$ are all left arbitrary at this stage.

\textcolor{black}{On the other hand in} order to express results \textcolor{black}{hereafter} in a convenient form, the following \textcolor{black}{additional}
notations are useful,
\begin{eqnarray}
X_2(\cos\theta) &=& (\sigma^2+\delta^2)-2(\sigma^2+\delta^2-1)\,\cos^2\theta, \nonumber \\
X_3(\cos\theta) &=& (\sigma^2+\delta^2-2)\,\sin^2\theta=(\sigma^2+\delta^2-2)(1-\cos^2\theta), \nonumber \\
X_4(\cos\theta) &=& \left[(\sigma^2 \delta^2 +1)-(2\sigma^4+2\delta^4+11\sigma^2 \delta^2 -2 \sigma^2 - 2\delta^2-1)\,\cos^2\theta\right]\,\sin^2\theta.
\end{eqnarray}

Proceeding as indicated \textcolor{black}{by using (\ref{eq:A}) (but leaving aside here its lengthy calculational though straightforward details)},
a careful evaluation of all contributions to the \textcolor{black}{asympotic} expansion for the potential $A$ (which scales as $r^2$) finds,
\begin{eqnarray}
{\rm Re}\,\left[\frac{1}{r^2}A\right] &=& -8m^6(\sigma^2-\delta^2)\sigma^2 \delta^2
\left[1+X_2\cdot u^2\,-\,X_4\cdot u^4\,+\,{\cal O}(u^6)\right], \nonumber \\
{\rm Im}\,\left[\frac{1}{r^2}A\right] &=& -32 m^6\cdot\frac{k\nu}{m^2}\cdot(\sigma^2-\delta^2)\sigma^2 \delta^2\,
\sin^2\theta\,\cos\theta\cdot u^3\,+\,{\cal O}(u^5),
\label{eq:A1A2}
\end{eqnarray}
and likewise \textcolor{black}{by using (\ref{eq:B})} for the potential $B$ (which scales as $r$),
\begin{eqnarray}
{\rm Re}\,\left[\frac{1}{r}\cdot\frac{1}{r}B\right] &=& -16m^6 (\sigma^2-\delta^2)\sigma^2 \delta^2\left[u\,+\,\frac{1}{2}X_3\cdot u^3\,+\,{\cal O}(u^5)\right], \nonumber \\
{\rm Im}\,\left[\frac{1}{r}\cdot\frac{1}{r}B\right] &=& -16m^6\cdot\frac{k\nu}{m^2}\cdot(\sigma^2-\delta^2)\sigma^2 \delta^2\,
\left[\cos\theta\cdot u^2\,-\,\frac{1}{2}X_3\cos\theta\cdot u^4\,+\,{\cal O}(u^6)\right],
\label{eq:B1B2}
\end{eqnarray}
as well as, finally, for the combination $A+B=r^2(A/r^2+(B/r)/r)$,
\begin{eqnarray}
{\rm Re}\,\left[\frac{1}{r^2}(A+B)\right]\!\!\! &=& \!\!\! -8m^6(\sigma^2-\delta^2)\sigma^2 \delta^2
\left[1+ 2\cdot u + X_2\cdot u^2\,+\,X_3\cdot u^3\,-\,X_4\cdot u^4\,+\,{\cal O}(u^5)\right], \nonumber \\
{\rm Im}\,\left[\frac{1}{r^2}(A+B)\right]\!\!\! &=& \!\!\! -8 m^6\cdot\frac{k\nu}{m^2}\cdot(\sigma^2-\delta^2)\sigma^2 \delta^2\, \nonumber \\
&& \times \left[2\cos\theta\cdot u^2\,+\,4\sin^2\theta\,\cos\theta\cdot u^3\,-\,X_3\cos\theta\cdot u^4\,+\,{\cal O}(u^5)\right].
\end{eqnarray}
Note how the imaginary parts of $A$ and $B$ are directly linear in the NUT charge through the combination $k\nu/m^2=2\xi$.

Given the definition of these quantities in terms of the potentials $A$ and $B$ as expressed in (\ref{eq:AB}),
these results allow for the asymptotic representation of the contributions in $f$ and $f^{-1}e^{2\gamma}$ to the metric components \textcolor{black}{in (\ref{eq:metric})}. 
Here only the detailed expression for $f$ is listed (\textcolor{black}{since} that for $f^{-1}e^{2\gamma}$ is not required in the remainder of the present analysis), with,
\begin{eqnarray}
f &=& 1-4\cdot u+8\cdot u^2+2\left[(\sigma^2+\delta^2-6)-3\left(\sigma^2+\delta^2-\frac{2}{3}\right)\,\cos^2\theta\right]\cdot u^3 \nonumber \\
&& -8\left[(\sigma^2+\delta^2-2)-3\left(\sigma^2+\delta^2-\frac{2}{3}-\frac{1}{3}\left(\frac{k\nu}{m}\right)^2\right)\,\cos^2\theta\right]\cdot u^4\,+\,{\cal O}(u^5),
\label{eq:f}
\end{eqnarray}
while one has for $e^{2\gamma}$, to lowest \textcolor{black}{nontrivial} order for the sake of illustration,
\begin{equation}
e^{2\gamma}=1-4\sin^2\theta\cdot u^2\,+\,{\cal O}(u^4).
\end{equation}
Note well that these results manifest explicitly the asymptotically flat character of the gravimagnetic dipole space-time metric,
for all values of $(m,k,\nu)$ such that $\alpha_\pm$ take real values.

The asymptotic expansion of $\omega(\rho,z)$ requires the expansion of the potentials $G_1$ and $G_2$, which scale as $r$ and $r^2$,
respectively. However since the angular line element is $\rho\,d\varphi$ \textcolor{black}{while having in mind the use of this expansion for the
evaluation of the contributions to the velocity curve of the gravito-electromagnetic fields of Sect.\ref{Sect4}, it turns out, as established in Sect.\ref{Sect4},
that} the required expansion needs to be exact only up to order $u^2$ inclusive,
since $\omega d\varphi=(\omega/\rho)\,\rho\,d\varphi$. Furthermore note that $\omega$ is necessarily linear in the
NUT charge through $k\nu/m^2=2\xi$, because it is given by a pure imaginary contribution (see (\ref{eq:AB})).

These expansions in $u$ for $G_1$ and $G_2$ are not as compact and streamlined as those for $A$ and $B$ above. One finds,
\textcolor{black}{using this time (\ref{eq:G2}) and (\ref{eq:G1}),}
\begin{equation}
{\rm Re}\,\left[\frac{1}{r}\cdot\frac{1}{r} G_1\right]=-16m^7(\sigma^2-\delta^2)\sigma^2\delta^2\cdot u\,-\,
8m^7(\sigma^2-\delta^2)\sigma^2\delta^2(\sigma^2+\delta^2-2)\,\cos\theta\cdot u^2 \,+\,{\cal O}(u^3),
\label{eq:G11}
\end{equation}
\begin{equation}
{\rm Im}\,\left[\frac{1}{r}\cdot\frac{1}{r} G_1\right]=-16m^7(\sigma^2-\delta^2)\sigma^2\delta^2\cdot\frac{k\nu}{m^2}\cdot u\,+\,
8m^7(\sigma^2-\delta^2)\sigma^2\delta^2\cdot\frac{2k\nu}{m^2}\,\cos\theta\cdot u^2\,+\,{\cal O}(u^3),
\label{eq:G12}
\end{equation}
as well as,
\begin{eqnarray}
{\rm Re}\,\left[\frac{1}{r^2} G_2\right] &=&16 m^7 (\sigma^2-\delta^2)\sigma^2 \delta^2\,\cos\theta\cdot u \nonumber \\
&& - 4 m^7(\sigma^2-\delta^2)(\sigma^2\delta^2+1)[1+(\sigma^2+\delta^2)\sin^2\theta\cdot u^2] \nonumber \\
&& + 4m(\sigma^2-\delta^2)[d^2k^2-(2k^2-m^2)(m^2+\nu^2)^2+k^2\nu^4]\,\cos^2\theta\cdot u^2 \nonumber \\
&& - 4 m^7 (\sigma^2-\delta^2)(\sigma^2\delta^2 -1)[1+(\sigma^2+\delta^2)(\sin^2\theta - \cos^2\theta)\cdot u^2] \nonumber \\
&& + 8m^7(\sigma^2-\delta^2)\sigma^2\delta^2\,\cos\theta\,[1+\frac{1}{2}(\sigma^2+\delta^2-2)\sin^2\theta\cdot u^2] \,+\,{\cal O}(u^3),
\label{eq:G21}
\end{eqnarray}
\begin{eqnarray}
{\rm Im}\,\left[\frac{1}{r^2} G_2\right] &=& 16 m^7(\sigma^2-\delta^2)\sigma^2\delta^2\cdot\frac{k\nu}{m^2}\,\cos^2\theta\cdot u^2 \nonumber \\
&& + 8m^7 (\sigma^2-\delta^2) \sigma^2 \delta^2\cdot\frac{k\nu}{m^2}\,\cos^2\theta\cdot u \,+\,{\cal O}(u^3).
\label{eq:G22}
\end{eqnarray}

The careful expansion of the combinations of products of real and
imaginary parts of $(G_1+G_2)$ and $(A+B)$ \textcolor{black}{as they are involved in the expression for $\omega$ in (\ref{eq:AB})} then establishes that,
\textcolor{black}{to lowest order,}
\begin{equation}
\omega=-4m\cdot\frac{k\nu}{m^2}\,\left[\sin^2\theta\cdot u\,-\,4\cos^2\theta\cdot u^2\right]\,+\,{\cal O}(u^3),
\label{eq:omega}
\end{equation}
as well as,
\begin{equation}
f\,\omega=-4m\cdot\frac{k\nu}{m^2}\,\left[\sin^2\theta\cdot u\,-\,4\cdot u^2\right]\,+\,{\cal O}(u^3),\qquad
f\,\omega^2=16m^2\cdot\left(\frac{k\nu}{m^2}\right)^2\,\sin^4\theta\cdot u^2\,+\,{\cal O}(u^3).
\label{eq:omega2}
\end{equation}

\section{The Gravito-Electromagnetic Approximation}
\label{Sect4}

In order to identify the gravito-electric and -magnetic fields relevant to the Lorentz-like equation of motion
as the gravito-electromagnetic approximation for the geodesic equation of a massive test point particle in the nonrelativistic
regime, we follow the careful and detailed discussion of Ref.\cite{Williams} using the same notations except for the additional
use of natural units such that $c=1$ and $G=1$.

The asymptotically flat space-time metric is thus considered in the following form,
\begin{equation}
g_{\mu\nu}=\eta_{\mu\nu} + h_{\mu\nu},\qquad h_{\mu\nu}\ll 1,
\end{equation}
with $\eta_{\mu\nu}$ being the usual Minkowski space-time metric (relative to cartesian coordinates $(t,x,y,z)$, say) with signature $(-+++)$.

The perturbations $h_{\mu\nu}$ then determine the gravito-electromagnetic potentials $(\phi,\vec{w},\Psi, s_{ij})$ through the following relations \cite{Williams},
\begin{equation}
h_{tt}=-2\phi,\qquad
h_{ti}=w^i,\quad i=1,2,3,
\end{equation}
\begin{equation}
\Psi=-\frac{1}{6}\delta^{ij}\,h_{ij},\qquad
s_{ij}=\frac{1}{2}h_{ij}+\Psi\,\delta_{ij}=\frac{1}{2}h_{ij}-\frac{1}{6}\delta_{ij}\,\delta^{k\ell}h_{k\ell},
\end{equation}
$s_{ij}$ thus being symmetric and traceless, and representing the radiating part of the gravitational field.
The corresponding gravito-electromagnetic fields $(\vec{E}_g,\vec{B}_g,\vec{N}_g)$ are then given as \cite{Williams},
\begin{equation}
E^i_g=-\partial_i\phi-\partial_t\,w^i,\qquad
B^i_g=\epsilon^{ijk}\partial_j\,w^k,\qquad
N^i_g=-\partial_i\,\Psi,
\end{equation}
while the authors of Ref.\cite{Williams} emphasize the relevance and importance of also including the {\sl neutral field} $\vec{N}_g$
in the equations of motion that all these gravito-electromagnetic fields must obey (these equations of motion need not be considered here, see Ref.\cite{Williams}).

In terms of the nonrelativistic momentum $\vec{p}=\mu\,\dot{\vec{x}}$ of a test point particle of mass $\mu$, and expanded up to linear order
in its velocity, in such a nonrelativistic approximation the geodesic equation for such a particle then reduces to the following Lorentz-like form \cite{Williams},
\begin{equation}
\frac{d}{dt}\vec{p}=\mu\,\left(\vec{E}_g\,+\,\vec{v}\times\vec{B}_g\,+\,2\vec{v}\,\partial_t\Psi\,-\,2v^j\,\partial_t\,s_{ij}\right)\,+\,{\cal O}(v^i\,v^j).
\end{equation}
In particular when the perturbation $h_{\mu\nu}$ is stationary, {\sl i.e.,} time independent, this equation is of the Lorentz form,
\begin{equation}
\frac{d}{dt}\vec{p}=\mu\,\left(\vec{E}_g\,+\,\vec{v}\times\vec{B}_g\right)\,+\,{\cal O}(v^i\,v^j),
\label{eq:Lorentz}
\end{equation}
thus requiring finally the knowledge only of the gravito-electromagnetic \textcolor{black}{scalar and vector} potentials $\phi$ and $\vec{w}$.

\subsection{The gravito-electromagnetic fields of the gravimagnetic dipole}

In order to apply the above discussion to the asymptotic expansion of the gravimagnetic dipole space-time metric,
let us also consider the cartesian coordinates $(x,y,z)$ related to the spherical and cylindrical Weyl coordinates used above, namely,
\begin{equation}
x=r\sin\theta\,\cos\varphi,\qquad
y=r\sin\theta\,\sin\varphi,\qquad
z=r\cos\theta,
\end{equation}
such that in particular $\rho^2=x^2+y^2=r^2\sin^2\theta=r^2(1-\cos^2\theta)$.

In the notations of Ref.\cite{Williams} \textcolor{black}{and given the parametrisation (\ref{eq:metric}) for the gravimagnetic dipole space-time metric},
one then has the following correspondencies,
\begin{equation}
h_{tt}=1-f,
\end{equation}
\begin{equation}
h_{tx}=-f\,\omega\,\frac{y}{x^2+y^2},\qquad
h_{ty}=f\,\omega\,\frac{x}{x^2+y^2},\qquad
h_{tz}=0,
\label{eq:wti}
\end{equation}
and finally,
\begin{eqnarray}
h_{xx} &=& \frac{1}{x^2+y^2}\left[x^2\left(f^{-1}e^{2\gamma}-1\right)+y^2\left(f^{-1}-f\frac{\omega^2}{x^2+y^2}-1\right)\right], \nonumber \\
h_{yy} &=& \frac{1}{x^2+y^2}\left[y^2\left(f^{-1}e^{2\gamma}-1\right)+x^2\left(f^{-1}-f\frac{\omega^2}{x^2+y^2}-1\right)\right], \nonumber \\
h_{zz} &=& \left(f^{-1} e^{2\gamma} -1 \right), \nonumber \\
h_{xy} &=& \frac{xy}{x^2+y^2}\left[\left(f^{-1}e^{2\gamma}-1\right)-\left(f^{-1}-f\frac{\omega^2}{x^2+y^2}-1\right)\right], \nonumber \\
h_{xz} &=& 0, \nonumber \\
h_{yz} &=& 0.
\end{eqnarray}
As noted above, in order to apply the Lorentz-like equation of motion (\ref{eq:Lorentz}) only the 3-scalar and 3-vector fields
$h_{tt}=-2\phi$ and $h_{ti}=w^i$ are required.

Since $h_{tt}=-2\phi=1-f$ \textcolor{black}{and with $f$ as given in (\ref{eq:f})}, the gravito-electric scalar potential is thus, up to order $1/r^4$ inclusive,
\begin{equation}
\phi(r,\cos\theta)=\frac{1}{2}(f-1)=-\frac{2m}{r}+4\left(\frac{m}{r}\right)^2+\frac{1}{2}f_3(\cos\theta)\left(\frac{m}{r}\right)^3
+\frac{1}{2}f_4(\cos\theta)\left(\frac{m}{r}\right)^4+\cdots,
\end{equation}
with \textcolor{black}{the quantities $f_3(\cos\theta)$ and $f_4(\cos\theta)$ defined by},
\begin{equation}
f_3(\cos\theta)=8\left[\left(\alpha-1\right)-\left(3\alpha + 1\right)\cos^2\theta\right],\qquad
f_4(\cos\theta)=-32\left[\alpha-\left(3\alpha+1-\xi^2\right)\cos^2\theta\right],
\end{equation}
\textcolor{black}{while the relations} $\alpha=(\sigma^2+\delta^2-2)/4$ and $\xi=k|\nu|/(2m^2)$ have now been substituted \textcolor{black}{as well}.
Up to that same relative order in $1/r$, the radial component of the corresponding gravito-electric field $\vec{E}_g=-\vec{\nabla}\phi$ is then given by,
\begin{equation}
\left(\vec{E}_g(r,\cos\theta)\right)_r=-\frac{\partial\phi(r,\cos\theta)}{\partial r}=-\frac{2m}{r^2}+\frac{8m^2}{r^3}+\frac{3}{2}f_3(\cos\theta)\frac{m^3}{r^4}
+2f_4(\cos\theta)\frac{m^4}{r^5}+\cdots,
\end{equation}
with in particular in the $z=0$ or $\cos\theta=0$ plane,
\begin{equation}
\left(\vec{E}_g(r,\cos\theta=0)\right)_r=-\frac{2m}{r^2}+\frac{8m^2}{r^3}+12(\alpha-1)\frac{m^3}{r^4} - 64\alpha \frac{m^4}{r^5}+\cdots,
\end{equation}
as well as,
\begin{equation}
\phi(r,\cos\theta=0)=-\frac{2m}{r}+\frac{4m^2}{r^2}+4(\alpha-1)\frac{m^3}{r^3}-16\alpha\frac{m^4}{r^4}+\cdots ,
\end{equation}
so  that
\begin{equation}
r\frac{\partial \phi(r,\cos\theta)}{\partial r}_{|_{\cos\theta=0}}=\frac{2m}{r}-8\left(\frac{m}{r}\right)^2-12(\alpha-1)\left(\frac{m}{r}\right)^3+64\alpha\left(\frac{m}{r}\right)^4
+\cdots.
\end{equation}
Of course one recognises in the first contribution in $-2m/r$ (for $\phi$) and $-2m/r^2\,\hat{e}_r$ (for $\vec{E}_g$)
the usual newtonian gravitational potential and field, namely $-2Gm/r$ and $-2Gm/r^2\,\hat{e}_r$, respectively,
of a point mass of value $M=2m$ (in the units $c=1$ and $G=1$), as ought to be the case.

Higher order corrections in powers of $1/r$ however, are represented by an alternating sign series, in fact two \textcolor{black}{such series intertwined}.
As discussed in the Appendix, in order to improve the small $r$ behaviour of such expansions to better represent the actual solution for the 3-scalar
potential one may apply a resummation of the series in the form of (\ref{eq:modulation}) with a correction or modulation factor involving a particular
distance scale factor $\lambda$ and a specific power $\beta>0$, such that,
\begin{equation}
\frac{1}{x}-\beta\frac{\lambda^2}{x^3}+\cdots\simeq \frac{1}{x}\cdot\frac{1}{(1+\lambda^2/x^2)^\beta},
\label{eq:beta}
\end{equation}
where $x=r/m$ (as well as the use of natural units such that $G=1$ and $c=1$). As illustrated in the Appendix, the case of a continuously distributed axisymmetric
mass distribution is expected to be well represented when using a power $\beta=1/2$ (as opposed to $\beta=3/2$ in the case of two point masses).
Given the presence of the Misner string linking the two black holes of the gravimagnetic dipole, $\beta=1/2$ remains the preferred value as well
to be implemented hereafter.

By using this observation as a rationale for a reliable resummation of the above series representation, one is readily led to the following organisation of terms
involving two intertwined alternating sign series, indeed with $x=r/m$,
\begin{eqnarray}
r\frac{\partial \phi(r,\cos\theta)}{\partial r}_{|_{\cos\theta=0}} &=& \frac{2}{x}-\frac{8}{x^2}-\frac{12(\alpha-1)}{x^3}+\frac{64\alpha}{x^4}+\cdots \nonumber \\
&=& \frac{2}{x}\left[1-\frac{4}{x}-\frac{6(\alpha-1)}{x^2}+\frac{32\alpha}{x^3}+\cdots\right] \nonumber \\
&=& \frac{2}{x}\left[\left(1-\frac{6(\alpha-1)}{x^2}+\cdots\right)\,-\,\frac{4}{x}\left(1-\frac{8\alpha}{x^2}+\cdots\right)\right] \nonumber \\
&\simeq& \frac{2}{x}\left[\left(1+12(\alpha-1)/x^2\right)^{-1/2}-\frac{4}{x}\left(1+16\alpha/x^2\right)^{-1/2}\right] \nonumber \\
&\simeq& \frac{2}{x}\cdot\frac{1}{\sqrt{1+12(\alpha-1)/x^2}}\cdot\left[1-\frac{4}{x}\cdot\sqrt{\frac{1+12(\alpha-1)/x^2}{1+16\alpha/x^2}}\right].
\label{eq:resummation}
\end{eqnarray}
Hence it is in this latter form that the weak field approximation for the gravito-electric scalar potential and its contribution to the rotation curve\footnote{Were it not
for the gravito-magnetic field contribution to the final rotation curve to be addressed presently, as a matter of fact the expression in (\ref{eq:resummation})
determines already the quantity $v^2(r)/c^2$, namely the square of the velocity rotation curve in units of $c$.}
is to be used hereafter, as it ought to much better represent the radial variation of that quantity than the above finite order series approximation
with its wild and unphysical fluctuations as $r$ decreases as a consequence of its alternating signs.

Turning now to the gravito-magnetic field \textcolor{black}{$w^i=h_{ti}$ with its components given in (\ref{eq:wti}) in which the substitution of
the result (\ref{eq:omega2}) for $f\omega$ is now to be effected,}
to lowest order in $1/r$ \textcolor{black}{these components} of the gravi-magnetic 3-vector potential read,
\begin{eqnarray}
w^1=w_x=h_{tx} &=& 4\cdot\frac{k\nu}{m^2}\,\sin\theta\,\sin\varphi\cdot u^2, \nonumber \\
w^2=w_y=h_{ty} &=& -4\cdot\frac{k\nu}{m^2}\,\sin\theta\,\cos\varphi\cdot u^2, \nonumber \\
w^3=w_z=h_{tz} &=& 0,
\end{eqnarray}
which are thus already of order $1/r^2$. In cartesian coordinates the gravito-magnetic vector potential is thus such that
\begin{equation}
w^1=4k\nu\,\frac{y}{(x^2+y^2+z^2)^{3/2}},\qquad
w^2=-4k\nu\,\frac{x}{(x^2+y^2+z^2)^{3/2}},\qquad
w^3=0.
\end{equation}
One recognises in this field the vector potential of a magnetic dipole with its magnetic moment aligned with the $z$-axis
and a magnetic moment value (proportional to) $(-4k\nu)$.

Consequently the gravito-magnetic field is given as,
\begin{eqnarray}
B^1_g=\left(\vec{\nabla}\times\vec{w}\right)^1 &=& -12\,k\nu\,\frac{xz}{(x^2+y^2+z^2)^{5/2}}, \nonumber \\
B^2_g=\left(\vec{\nabla}\times\vec{w}\right)^2 &=& -12\,k\nu\,\frac{yz}{(x^2+y^2+z^2)^{5/2}}, \nonumber \\
B^3_g=\left(\vec{\nabla}\times\vec{w}\right)^3 &=& -4\,k\nu\,\frac{2z^2-(x^2+y^2)}{(x^2+y^2+z^2)^{5/2}}.
\end{eqnarray}
In particular in the $z=0$ or $\cos\theta=0$ plane, one has,
\begin{equation}
\vec{B}_g(x,y,z=0)=\vec{\nabla}\times\vec{w}=\frac{4k\nu}{r^3}\,\hat{e}_z,
\end{equation}
thereby determining an additional contribution to the velocity rotation curve through the Lorentz-like force equation of motion
(\ref{eq:Lorentz}).

\subsection{The rotation curve for circular trajectories at $z=0$}

In the nonrelativistic limit and given the gravimagnetic dipole space-time metric in the \textcolor{black}{weak field}
gravito-electromagnetic approximation under consideration,
the Lorentz-like equation of motion (\ref{eq:Lorentz}) thus reduces to,
\begin{equation}
\frac{d}{dt}\vec{v}=\vec{E}_g+\vec{v}\times\vec{B}_g.
\end{equation}
When restricting to a circular trajectory of radius $r$ in the plane $z=0$ or $\cos\theta=0$ one has,
\begin{equation}
\vec{v}=v(r)\,\hat{e}_\varphi,\qquad
\vec{v}\times\vec{B}_g(x,y,z=0)=\frac{4k\nu}{r^3}\,v(r)\,\hat{e}_r,
\end{equation}
where $v(r)$ is the tangential component of the velocity vector in the $\varphi$ direction,
which could be of either sign.
Under these specific circumstances the equation of motion reduces to
\begin{equation}
\frac{v^2(r)}{r}=\frac{\partial\phi(r,\cos\theta)}{\partial r}_{|{\cos\theta=0}}\,-\,\frac{4k\nu}{r^3}\,v(r).
\end{equation}

The rotation curve $v(r)$ is thus determined from the quadratic equation,
\begin{equation}
v^2(r)+\frac{4k\nu}{r^2}\,v(r) - r\frac{\partial\phi(r,\cos\theta)}{\partial r}_{|{\cos\theta=0}}=0,
\end{equation}
leading to the following two solutions, in units of $c$,
\begin{equation}
v_+(r)=\sqrt{ r\frac{\partial\phi(r,\cos\theta)}{\partial r}_{|{\cos\theta=0}}+\frac{4k^2\nu^2}{r^4}}\,-\,\frac{2k\nu}{r^2},
\end{equation}
\begin{equation}
v_-(r)=-\sqrt{ r\frac{\partial\phi(r,\cos\theta)}{\partial r}_{|{\cos\theta=0}}+\frac{4k^2\nu^2}{r^4}}\,-\,\frac{2k\nu}{r^2}.
\end{equation}
In terms of the quantities $x=r/m$, $\xi=k|\nu|/(2m)$ and $\alpha=(\sigma^2+\delta^2-2)/4$, and using for the gravito-electric scalar field
contribution the resummed expression in the form of (\ref{eq:resummation}), one has,
\begin{equation}
v_\pm(r)=\pm\sqrt{ \frac{2}{x}\cdot\frac{1}{\sqrt{1+12(\alpha-1)/x^2}}\cdot\left(1-\frac{4}{x}\cdot\sqrt{\frac{1+12(\alpha-1)/x^2}{1+16\alpha/x^2}}\right)
+\frac{16\xi^2}{x^4}}\,-\,\frac{4\xi}{x^2},
\label{eq:vpm}
\end{equation}
or equivalently,
\begin{equation}
v_\pm(r)=\pm\sqrt{\frac{2}{x}}\cdot\frac{1}{\left(1+12(\alpha-1)/x^2\right)^{1/4}}\cdot
\sqrt{1-\frac{4}{x}\sqrt{\frac{1+12(\alpha-1)/x^2}{1+16\alpha/x^2}}\,+\,\frac{8\xi^2}{x^3}\sqrt{1+\frac{12(\alpha-1)}{x^2}}}\,-\,\frac{4\xi}{x^2},
\label{eq:vpm2}
\end{equation}
which clearly displays the expected leading asymptotic behaviour in $\sqrt{2m/r}$.\footnote{In case another value than $\beta=1/2$ is preferred
in the implementation of resummations in the form of (\ref{eq:beta}), say $\beta=n/2$ for some $n>0$, the corresponding results for $v_\pm(r)$ read,
still with the same leading asymptotic behaviour,\\
$
v_\pm(r)=\pm\sqrt{\frac{2}{x}}\cdot\left(1+12(\alpha-1)/x^2\right)^{-n/4}\cdot
\sqrt{1-\frac{4}{x}\left(\frac{1+12(\alpha-1)/x^2}{1+16\alpha/x^2}\right)^{n/2}\,+\,\frac{8\xi^2}{x^3}\left(1+\frac{12(\alpha-1)}{x^2}\right)^{n/2}}\,-\,\frac{4\xi}{x^2}
$.}
 A few more comments are in order as well.
 
Since the lowest order contribution in the $1/r$ expansion to the rotation curve $v^2(r)/c^2$ of the gravito-magnetic correction which is being added
to  the gravito-electric scalar field contribution, namely $r\partial\phi(r,\cos\theta=0)/\partial r$, is of order $1/r^4$, the latter contribution in $\phi$
had to be established at least to the same $1/r^4$ order inclusive, as is indeed done above.

On the other hand, note that in order for these expressions to be physically reliable one must be in a regime
such that $|v_\pm(r)| \ll 1$, since they are established within a nonrelativistic approximation.

Obviously the presence of the NUT gravimagnetic dipole with nonvanishing NUT charges $\pm\nu$ breaks rotational (and time reversal) invariance
by distinguishing the two spatial chiralities, hence the two rotation directions relative to the $z$ symmetry axis,
with different values for $v_+(r)$ and $v_-(r)$ when $\xi\ne 0$ or $\nu\ne 0$.
However, since the gravimagnetic dipole space-time metric is asymptotically flat, the asymptotic behaviour of these rotation curves displays the usual
Keplerian profile in $\sqrt{2m/r}$ for sufficiently large values of $r/m$, irrespective of the rotation direction, of the value of the NUT charges,
and of relativistic corrections.

More specifically now, and first for the purpose of comparison with the newtonian result for the cylindrical uniform mass distribution of total length $2L$
as discussed in the Appendix, let us consider the above expressions for $v_\pm(r)$ when $\xi=0$ or $\nu=0$, in which case,
\begin{equation}
\xi=0:\quad
v_\pm(r)=\pm\sqrt{\frac{2}{x}}\cdot\frac{1}{\left(1+12(\alpha-1)/x^2\right)^{1/4}}\cdot
\sqrt{1-\frac{4}{x}\sqrt{\frac{1+12(\alpha-1)/x^2}{1+16\alpha/x^2}}},
\end{equation}
the parameter $\alpha$ then being given as $\alpha=(k^2/m^2-1)/4$, $2k$ being the total distance between the two black holes (in units of $m$).
While rotational invariance around the symmetry axis $z$ is then restored with in particular $|v_+(r)|=|v_-(r)|$,
the newtonian result is reproduced by the first two factors in the r.h.s.~of the above expression when using
the identification $L=m\sqrt{3(k^2/m^2-5)}$ (in units of $m$). Given this comparison, the third factor in the above expression ought thus to correspond
to a relativistic correction contribution that reduces to unity in the limit $c\rightarrow\infty$, as indeed $1/x=Gm/(r c^2)$.

As mentioned in the Appendix, the second factor $(1+12(\alpha-1)/x^2)^{-1/4}$ which modulates the overall asymptotic behaviour in $\sqrt{2/x}$
corrects the latter sufficiently already to lead to a reasonably well flattened out rotation curve up to distances $r\simeq L$. This appealing feature
is thus improved even further through the above extra third factor and second modulation contribution in the character of a relativistic correction,
on account of the negative sign for the contribution $-4/x\cdot\sqrt{(1+12(\alpha-1)/x^2)/(1+16\alpha/x^2)}$ inside that last square root factor.

Such a flattening out of the rotation curve is enhanced even further still, specifically in the case of the solution $v_+(r)$, when the gravito-magnetic field correction
proportional to a nonvanishing NUT charge, $\nu\ne 0$ or $\xi\ne 0$ is included as well, because of the competition between the positive overall first square root term
on the r.h.s.~of (\ref{eq:vpm}) and the subtracted contribution in $4\xi/x^2>0$.\footnote{We recall that without loss of generality $\nu$ is assumed
to be positive throughout.} This feature is made more explicit by expressing that solution in the form of,
\begin{eqnarray}
v_+(r) &=& \sqrt{\frac{2}{x}}\cdot\frac{1}{\left(1+12(\alpha-1)/x^2\right)^{1/4}}\times \nonumber \\
&&\quad \times \frac
{1-\frac{4}{x}\sqrt{\frac{1+12(\alpha-1)/x^2}{1+16\alpha/x^2}}}
{\sqrt{1-\frac{4}{x}\sqrt{\frac{1+12(\alpha-1)/x^2}{1+16\alpha/x^2}}+\frac{8\xi^2}{x^3}\sqrt{1+\frac{12(\alpha-1)}{x^2}}}
\,+\,\frac{4\xi}{x^2}\sqrt{\frac{x}{2}}\left(1+\frac{12(\alpha-1)}{x^2}\right)^{1/4}},
\label{eq:vp}
\end{eqnarray}
which is also more amenable to numerical evaluations for large (dimensionless) values of $\alpha$ and $\xi$ and decreasing ones for $x$.

For a given value of $m$ which sets the scale of physical distances, this last approximation representation of the gravimagnetic dipole rotation curve $v_+(r)$
depends on the two independent parameters which label this class of solutions to the vacuum Einstein equations, namely
\begin{equation}
\alpha=\frac{1}{4}\left(\frac{k^2-\nu^2}{m^2}-1\right)\ge 0,\qquad
\xi=\frac{1}{2}\cdot\frac{k|\nu|}{m^2}\ge 0.
\end{equation}
Strictly speaking one ought to restrict to the tensionless limit of the Misner string, corresponding to those configurations
such that $\alpha=\xi\sqrt{\xi^2+1}$. In particular in that case and for large values of the then sole remaining free parameter $\xi\ge 0$,
it was pointed out before that one has \textcolor{black}{to leading order},
\begin{equation}
{\rm Tensionless\ limit}\textcolor{black}{,\quad \xi\rightarrow \infty}:\quad \alpha\simeq \xi^2,\qquad
\frac{1}{4}\cdot\frac{k^2}{m^2}\simeq \xi^2,\qquad
\frac{\nu^2}{m^2}\simeq 1.
\end{equation}
Hence even when the tensionless limit is not strictly enforced, but provided $\alpha$ and $\xi$ are both large and approximately such that
$\alpha\simeq \xi^2$, numerical values for $v_+(r)$ as obtained from (\ref{eq:vpm}) or (\ref{eq:vp}) would represent still the physically relevant tensionless situation.

Consequently under such circumstances, besides the value for $m$ which sets the physical length scale in terms of which the dimensionless distance
variable $x=r/m$ is measured (with $G=1$ and $c=1$), the only other two dimensionless variables which control the value of the dimensionless
velocity $v_+(r)$ (in units of $c$) is a combined choice for the values of $\alpha$ and $\xi$.

To assess the potential offered by the gravimagnetic dipole solution to account for the observed reality of flattened out rotation curves of spiral galaxies,
let us aim to reproduce the typical value $v_0$ for the plateau velocity of such curves (see for instance Fig.1 in Refs.\cite{Sofue1,Sofue2}),
which is of the order of $v_0\simeq 230$ km/s (and within the approximate range $150\lesssim v_0 \lesssim 300$ km/s) starting at the
visible edge of such galaxies, thus with the nonrelativistic value $v_0/c\simeq 7.7\times 10^{-4}$.

\begin{figure}
\begin{center}
\includegraphics[width=16cm]{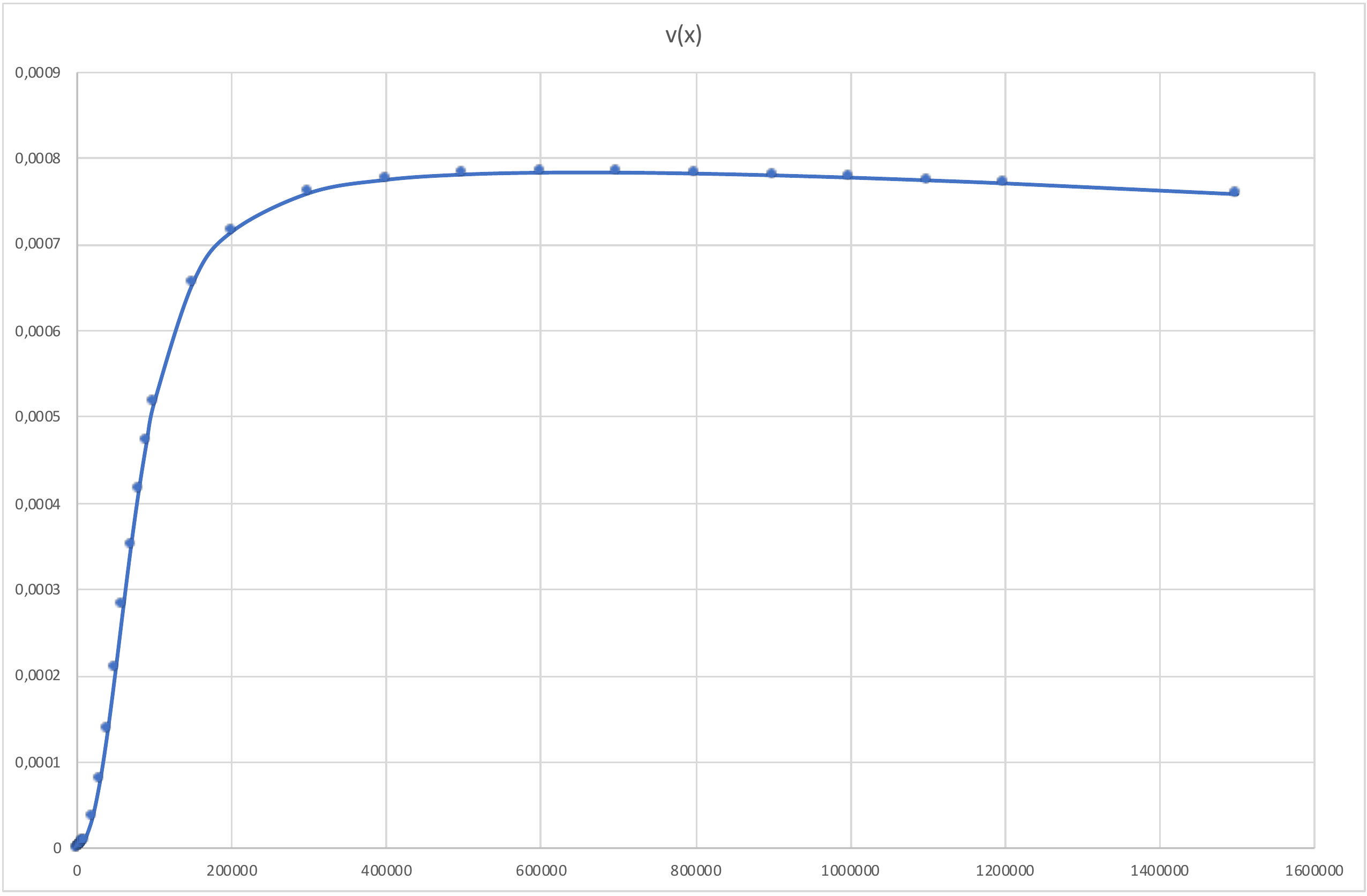}
\end{center}
\caption{The velocity rotation curve $v_+(r)$ (vertical axis, in units of $c$) given in Eqs.(\ref{eq:vpm},\ref{eq:vp}), as function of $x=r/m$ (horizontal axis, with
$0 \le x \le 1\,600\,000$, and units $G=1=c$), for the choice of dimensionless parameters $\xi=9\times 10^5$ and $\alpha=8.1\times 10^{11}$.}
\label{Fig1}
\end{figure}

Choosing for example
\begin{equation}
\xi=9\times 10^5,\qquad
\alpha=8.1\times 10^{11},
\end{equation}
indeed leads to a plateau value for the rotation curve $v_+(r)$ of
\begin{equation}
\frac{v_+(r)}{c}\simeq 7.8\times 10^{-4},
\end{equation}
with little variation in that velocity value within the distance range
\begin{equation}
400\,000 < x < 1\,500\,000.
\end{equation}
This configuration is illustrated in Fig.\ref{Fig1}. Correspondingly given these large values for $\xi$ and $\alpha$ one also has, still in units of $m$,
\begin{equation}
k\simeq 2\xi\,m,\qquad
\nu\simeq m,
\end{equation}
while the black hole horizon intercepts at $z=\alpha_\pm$ (or $z=-\alpha_\pm$) then take the values,
\begin{equation}
\frac{\alpha_+}{m}\simeq 1.8\times 10^6\simeq \frac{k}{m},\qquad
\frac{\alpha_-}{m}\simeq 1.8\times 10^6\simeq \frac{k}{m},\qquad
\frac{\alpha_+}{m}-\frac{\alpha_-}{m}=2\delta\simeq 2\sqrt{2}=2.828.
\end{equation}

The value of $m$ is determined from the physical distances characteristic of such plateaux in rotation curves.
Typical onsets of the plateau, around the visible edge of the spiral galaxy, lie in the range $r_0\simeq 5-15$ kpc, to correspond to $x\simeq 400\,000$.
Choosing values in that interval, and given the above values for $\alpha$ and $\xi$ with $\alpha\simeq \xi^2$,
the physical scales are then determined to be,
\begin{eqnarray}
r_0=5\ {\rm kpc}&:&\qquad m=2.6\times 10^{11}\,M_\odot,\qquad k=45\ {\rm kpc};  \nonumber \\
r_0=10\ {\rm kpc}&:&\qquad m=5.2\times 10^{11}\,M_\odot,\qquad k=90\ {\rm kpc};  \nonumber \\
r_0=15\ {\rm kpc}&:&\qquad m=7.8\times 10^{11}\,M_\odot,\qquad k=135\ {\rm kpc},
\end{eqnarray}
while for the black hole horizon intercepts on the $z$ axis one has for all these cases,
\begin{equation}
\alpha_+ - \alpha_- \simeq \frac{2\sqrt{2}}{k/m}\,k\simeq1.57\times 10^{-6}\,k
\end{equation}
(we recall that $1\ {\rm pc}=1\ {\rm parsec}\simeq 3.262$\ {\rm {light-years}}\ $\simeq 3.086\times 10^{13}\ {\rm km}$).
Such distance scale values for $k$ correspond to some fraction of the gross average distance between galaxies,
while those values for $m$ are comparable to the typical total visible mass of a spiral galaxy.

\textcolor{black}{
In view of the rather involved expression for $v_+(r)$ in (\ref{eq:vpm2}), simplified forms of it for various ranges in $r$ values may be useful.
Contrary to what may appear from Fig.\ref{Fig1}, the small $r$ limit of $v_+(r)$ does not go to $0$ for $r\rightarrow 0$, but rather reaches
that value for a nonvanishing value of $r>0$. However it should be kept in mind that (\ref{eq:vpm2}) itself is already a resummed approximation
to the asymptotic series in $1/r$  representing the velocity curve, which even though not beset by wild variations for smaller and smaller values
in $r$ is not to be relied upon too strictly for small values of $r$ where the gravitational field becomes strong and the weak field expansion no longer warranted.
Furthermore even if the small $r$ behaviour of $v_+(r)$ were to be reliably identified,
at some later stage one would still need to include on a case by case basis the contributions to specific galaxy velocity rotation curves
of their visible mass distributions.}

\textcolor{black}{
On the other hand, the situation regarding plateau values in $v_+(r)$ is a different matter. For $\xi$ and $\alpha\simeq \xi^2$ sufficiently large,
the plateau value in $v_+(r)$ corresponds a range of values in $r$ such that $x$ lies around $x\simeq\xi$. Based on (\ref{eq:vpm2})
one then establishes the following leading approximation for the plateau value of the velocity curve,
\begin{equation}
v_+^{\rm plateau} \simeq \sqrt{\frac{2}{\sqrt{13}}}\,\frac{1}{\sqrt{\xi}},\qquad
\sqrt{\frac{2}{\sqrt{13}}}\simeq 0.754.
\end{equation}}

\textcolor{black}{
The range of typical plateau values quoted above, namely $150\lesssim v_0 \lesssim 300$ km/s (see for instance Fig.1 in Refs.\cite{Sofue1,Sofue2}),
then corresponds to the following range of values in the parameter $\xi\simeq\sqrt{\alpha}$,
\begin{center}
\begin{tabular}{l l l}
$v_0=150\ {\rm km/s}$: &  $v_0/c =5\times 10^{-4}$, & \quad $\xi=2.22\times 10^6$;   \\
$v_0=200\ {\rm km/s}$: &  $v_0/c =6.67\times 10^{-4}$, & \quad $\xi=1.28\times 10^6$;   \\
$v_0=230\ {\rm km/s}$: &  $v_0/c =7.67\times 10^{-4}$, & \quad $\xi=9.44\times 10^5$;   \\
$v_0=234\ {\rm km/s}$: &  $v_0/c =7.8\times 10^{-4}$, & \quad $\xi=9.12\times 10^5$;   \\
$v_0=250\ {\rm km/s}$: &  $v_0/c =8.33\times 10^{-4}$, & \quad $\xi=7.99\times 10^5$;   \\
$v_0=300\ {\rm km/s}$: &  $v_0/c =1\times 10^{-3}$, & \quad $\xi=5.55\times 10^5$.
\end{tabular}
\end{center}
For such large values of $\xi$, the gravimagnetic dipole parameters ($k,\nu$) are given as $k\simeq 2\xi\,m$ and $\nu\simeq m$.}

Of course such values represent gross approximate \textcolor{black}{estimates} of these physical quantities, based on the weak field
asymptotic expansion of the gravimagnetic dipole space-time metric to allow for a first exploration through a gravito-electromagnetic assessment
of the rotation curve of such a system, without having included any other possible visible mass distribution\textcolor{black}{, say} in the case of a spiral galaxy.
The ``dark matter'' of such a system that may thus possess flattened rotation curves in some range for its parameters,
consists of two rotating massive black holes of equal masses carrying as well large and opposite NUT (or gravi-magnetic) charges,
and positioned along its rotation axis at a distance much larger than the visible spatial extent of some associated spiral galaxy.
While quite obviously, for a more realistic comparison, the contributions of all the visible mass that composes such a galaxy to its
rotation curve must still be included as well, in combination with those contributions stemming from the gravimagnetic dipole itself,
the latter of which have been assessed in the present work on the basis of a weak field gravito-electromagnetic approximation.

\section{Conclusions}
\label{Sect5}

Having assessed the overall profile of the velocity rotation curve of the gravimagnetic dipole within the limits
of weak field gravito-electromagnetism and nonrelativistic dynamics for a test point mass, this first analysis has
established that rotations curves which are sufficiently flattened out to become comparable to what is being observed for the
rotation curves of spiral galaxies, are indeed possible for a certain range of values for the parameters of these exact
solutions to the vacuum Einstein equations. Particularly noteworthy is the fact that the existence of nonvanishing
NUT charges for this system improves even further the situation, by increasing the width of the flattened out plateau in the rotation curve
with increasing values of the NUT charges.

In other words the flattened out profiles of the rotation curves of spiral galaxies may be a manifestation of some nonvanishing stationary axisymmetric
space-time curvature sourced by some energy-momentum distribution largely delocalised from the centre of the galaxy, and possibly carrying a NUT charge
as well, rather than being the effect of some dark matter particle halo harbouring a spiral galaxy.

Obviously however, for such a conclusion to be ascertained requires further thorough study of the physical potential offered by such a suggestion,
and in particular by the gravimagnetic dipole. On the one hand, a study of the geodesics of the gravimagnetic dipole space-time metric ought to be developed
without the approximations inherent to the methods used in the present work. And on the other hand, the contributions to its total velocity curve of the visible matter distributions comprising an actual spiral galaxy should also be included, as well as the possible presence of a massive black hole sitting at its
centre.\footnote{Note that there exist exact stationary axisymmetric $N$-black hole solutions to Einstein's vacuum solutions~\cite{ExactGR,Ruiz,Santos}.}
Both issues need to be pursued in order to try identify values of the relevant
parameters determining a particular gravimagnetic dipole when confronted to the physical reality of the flat rotation curve of any given spiral galaxy.

From a wider perspective, the suggestion entertained in the present work
certainly raises a series of other important questions. Would the values required for the parameters characteristic
of such solutions be physically realistic in view of the known proportion of missing mass or dark matter in the Universe as a whole? In the same way that it has been
proposed that primordial black holes may provide the necessary dark matter of the Universe, could it be that the Big Bang has witnessed the production of
such gravimagnetic dipoles carrying NUT charges, with a density decreasing with the expansion of the Universe, to eventually become the seeds for
structure and galaxy formation, and possibly account for at least part of the observed effects pointing to the reality of missing mass in the Universe
not only at the galactic scales, but at the astrophysical and cosmological scales as well? Which types of physical restrictions would thereby ensue, for instance
through limits on astrophysical lensing effects, possibly generated by the presence of Misner strings even in their tensionless limit? What is the stability of
such configurations? Could it be that missing mass observations are the manifestation of the existence of NUT matter in the Universe?
In view of the urgency and still total mystery of the dark matter conundrum of the Universe, investigations along such avenues deserve as well to be explored further.

\section*{Acknowledgements}

This work was largely completed while visiting the School of Physics at the University of Sydney (New South Wales, Australia) \textcolor{black}{in April 2022}.
The author is most grateful for the hospitality extended to him by Prof. Archil Kobakhidze and for extensive discussions
on the topic of this work, as well as Prof. C\'eline B{\oe}hm, \textcolor{black}{then} Head of the School of Physics. His work and visit are supported in part
by the {\sl Institut Interuniversitaire des Sciences Nucl\'eaires} (IISN, Belgium).

\section*{Appendix}
\label{Appendix}

Within the conceptual framework of newtonian gravity and its nonrelativistic mechanics, consider
the newtonian gravitational scalar potential sourced by a stationary axisymmetric mass distribution aligned
with the cartesian $z$ coordinate axis. Given a choice of spherical coordinates $(r,\theta,\varphi)$ relative to that symmetry axis,
the gravitational scalar potential $\phi(r,\cos\theta)$ is then $\varphi$-independent, with the associated newtonian gravitational field
determined from
\begin{equation}
\vec{g}(r,\cos\theta)=-\vec{\nabla}\phi(r,\cos\theta),\qquad
g_r(r,\cos\theta)=-\frac{\partial}{\partial r}\phi(r,\cos\theta),
\end{equation}
with in particular its radial component $g_r(r,\cos\theta)$.

As a consequence, for any circular trajectory with radius $r$ of a test point mass $\mu$ in the $z$-axis-perpendicular plane at $z=0$ or $\cos\theta=0$,
the norm $v(r)$ of its velocity is such that,
\begin{equation}
\mu\frac{v^2(r)}{r}=-\mu\,g_r(r,\cos\theta=0)=\mu\,r\frac{\partial}{\partial r}\phi(r,\cos\theta=0).
\end{equation}
The $r$-profile of that class of trajectories is thus characterised by the following ``rotation curve",
\begin{equation}
v(r)=\sqrt{r\frac{\partial\phi}{\partial r}(r,\cos\theta=0)},\quad
\frac{v(r)}{c}=\sqrt{\frac{1}{c^2}r\frac{\partial\phi}{\partial r}(r,\cos\theta=0)},\quad
\frac{v^2(r)}{c^2}=\frac{1}{c^2}\,r\frac{\partial \phi(r,\cos\theta)}{\partial r}_{|_{\cos\theta=0}},
\end{equation}
where in the last two expressions the rotation velocity is normalised to the speed of light in vacuum, $c$,
thereby providing the dimensionless quantity and profile $v(r)/c$ implied by the axisymmetric mass distribution under consideration.

Of course in the case of a point mass source of value $m$, one has the gravitational potential $\phi(r)=-Gm/r$, with $G$ being Newton's constant,
thus implying the well known velocity curve varying as $1/\sqrt{r}$, namely,
\begin{equation}
v(r)=\sqrt{\frac{Gm}{r}},\qquad
\frac{v(r)}{c}=\sqrt{\frac{G}{c^2}\cdot\frac{m}{r}},\qquad
\frac{v^2(r)}{c^2}=\frac{1}{x}.
\end{equation}
In this last expression for $v^2(r)/c^2$, the variable $x$ represents the distance $r$ measured in units of the gravitational length scale $Gm/c^2$ 
set by the mass value $m$, as defined by
\begin{equation}
x=\frac{r\,c^2}{Gm},\qquad r=\frac{Gm}{c^2}\,x.
\end{equation}
Given any mass distribution which remains confined to some spatial region of finite extent and volume, at sufficiently large distance away from it
and in a nonrelativistic limit its velocity curve will always display this unavoidable $1/\sqrt{x}$ behaviour as an overall multiplicative feature,
to be modulated by higher order correction effects in $1/r$ generated by the details of the spatial extent of the mass distribution,
as one moves closer to it.

The next simplest example is that of two identical point masses $m$ positioned symmetrically about $z=0$ and along the $z$-axis at a finite distance
from one another at the coordinate values $z=\pm \ell$ (with $\ell>0$). The gravitational scalar potential of this mass distribution is expressed as
\begin{equation}
\phi(r,\cos\theta)=-Gm\left(\frac{1}{\sqrt{r^2-2\ell r\cos\theta + \ell^2}}\,+\,\frac{1}{\sqrt{r^2+2\ell r\cos\theta + \ell^2}}\right).
\end{equation}
The corresponding rotation curve possesses the following radial dependency,
\begin{equation}
\frac{v(r)}{c}=\sqrt{\frac{G}{c^2}\cdot\frac{2m}{r}}\cdot\frac{1}{\left(1+\ell^2/r^2\right)^{3/4}},\qquad
\frac{v^2(r)}{c^2}=\frac{1}{c^2}\,r\frac{\partial\phi}{\partial r}_{|_{\cos\theta=0}}=\frac{2}{x}\cdot\frac{1}{(1+\lambda^2/x^2)^{3/2}},
\end{equation}
in terms of the dimensionless quantities $x$ and $\lambda$ defined by
\begin{equation}
x=\frac{r c^2}{Gm},\qquad
\lambda=\frac{\ell\,c^2}{Gm},
\end{equation}
which represent the distances $r$ and $\ell$ in units of $Gm/c^2$, respectively. In this case the overall large distance asymptotic behaviour in $2/x$ for $v^2(r)/c^2$
is modulated by the extra multiplicative factor $(1+\lambda^2/x^2)^{-3/2}$, with in particular the property that $\lim_{x\rightarrow 0}v(r)/c=0$
as must follow from the parity symmetry of the mass distribution under $z\rightarrow -z$.

More explicitly, the asymptotic expansion in $1/r$ of the rotation curve is provided by the series,
\begin{equation}
\frac{1}{c^2}\,r\frac{\partial\phi}{\partial r}_{|_{\cos\theta=0}}=\frac{2}{x}\cdot\left(1-\frac{3}{2}\frac{\lambda^2}{x^2}+\frac{15}{8}\frac{\lambda^4}{x^4}
-\frac{35}{16}\frac{\lambda^6}{x^6}+\cdots\right),
\end{equation}
so that,
\begin{equation}
\frac{v(r)}{c}=\sqrt{\frac{2}{x}-3\frac{\lambda^2}{x^3}+\frac{15}{4}\frac{\lambda^4}{x^5}-\frac{35}{8}\frac{\lambda^6}{x^7}+{\cal O}(x^{-9})}.
\end{equation}
Note however that due to the unavoidable alternating signs of this series, the proper behaviour of $v^2(r)/c^2$ around $x\simeq 0$ for $x$ sufficiently small
will never be reproduced to any degree of reliability when the series is considered only up to some finite order. Indeed any such finite order approximation
leads to a blow-up (to $+\infty$ or $-\infty$) in the value for $v^2(r)/c^2\ge 0$ around $x\simeq 0$, whereas one has $\lim_{x\rightarrow 0}v^2(r)/c^2=0$.
This unwelcome and unphysical behaviour may be remedied by resumming the series based on its first terms, in the form of
\begin{equation}
\frac{v^2(r)}{c^2}=\frac{2}{x}\cdot\left(1+\lambda^2/x^2\right)^{-\beta}\simeq\frac{2}{x}\cdot\left(1-\beta\,\frac{\lambda^2}{x^2}\,+\,\cdots\right),
\label{eq:modulation}
\end{equation}
for some value of the power $\beta>0$, and given a length scale $\lambda>0$ (in units of $Gm/c^2$). The above result shows that the value $\beta=3/2$
applies to a system with two well localised and separated point masses positioned along the $z$-symmetry axis.

In contradistinction, let us now address the situation of an axisymmetric mass distribution which extends continuously along the $z$ axis between two extreme values
for $z$. As the simplest such choice \cite{Govaerts}, let us consider a cylinder of uniform mass density $\rho$, of radius $a$ and of total length (or height) $2L>0$
along $z$, positioned symmetrically around $z=0$ to occupy the interval $-L\le z\le L$. It then follows that the corresponding exact rotation curve is provided
by the integral representation \cite{Govaerts},
\begin{equation}
r\ne a:\quad
\frac{v^2(r)}{c^2}=\frac{G}{c^2}\cdot 2m\cdot\frac{1}{\pi a}\cdot\frac{r}{L}\int_0^{2\pi}d\phi\,\cos\phi\,{\rm asinh}\,\left(\frac{L}{\sqrt{r^2+a^2-2ar\cos\phi}}\right),
\end{equation}
where $m=\pi a^2\,L\,\rho$ denotes half of the total mass $2m$ of the cylinder.

It turns out that to an excellent degree of approximation and this already for values 
of $r>a$ that differ very little from $a$ (in relative terms), the above expression aligns numerically extremely well \cite{Govaerts} with the following behaviour in $r$,
which is indeed perfectly independent from the cylinder radius value $a$ (when expressed in terms of its total mass $2m$),
\begin{equation}
r>a:\quad \frac{v(r)}{c}=\sqrt{\frac{G}{c^2}\cdot\frac{2m}{L}}\cdot\frac{1}{\left(1+r^2/L^2\right)^{1/4}},\qquad
\frac{v^2(r)}{c^2}=\frac{G}{c^2}\cdot\frac{2m}{r}\cdot\frac{1}{\left(1+L^2/r^2\right)^{1/2}},
\end{equation}
namely,
\begin{equation}
\frac{v^2(r)}{c^2}=\frac{2}{x}\cdot\frac{1}{\left(1+\lambda^2/x^2\right)^{1/2}},
\label{eq:cylinder}
\end{equation}
with once again the dimensionless distance scale parameters,
\begin{equation}
x=\frac{r\,c^2}{Gm},\qquad
\lambda=\frac{L\,c^2}{G m}.
\end{equation}
The $1/r$ asymptotic expansion of this expression leads to the series representation
\begin{equation}
\frac{v^2(r)}{c^2}=\frac{2}{x}-\frac{\lambda^2}{x^3}+\frac{3}{4}\frac{\lambda^4}{x^5}-\frac{5}{8}\frac{\lambda^6}{x^7}+{\cal O}(x^{-9}),
\end{equation}
which again suffers from the unwelcome and unphysical behaviour noted above due to an alternating sign series.

However this may be remedied by resumming the series
based on its first terms, leading again to a modulation factor in the form of (\ref{eq:modulation}) but this time with the power $\beta=1/2$.\footnote{The
above result in (\ref{eq:cylinder}) implies a flattening out of the rotation curve $v(r)/c$, as compared to the behaviour in $\sqrt{2m/x}$, up to a distance scale
set by the value of $L$ representing the transverse extent of the mass distribution \cite{Govaerts}, thus without necessarily the need for some additional spherical
dark matter halo into which to immerse a spiral galaxy.}
This value for $\beta$ is thus characteristic of a continuously distributed mass density along the $z$ axis in an axisymmetric configuration,
to be contrasted with the above value $\beta=3/2$ that applies in the case of two localised point masses at a finite distance from one another.

\clearpage

\end{document}